# Quantifying the Ease of Scientific Discovery

Samuel Arbesman

*Department of Health Care Policy, Harvard Medical School, Boston, MA USA*

*Institute for Quantitative Social Science, Harvard University, Cambridge, MA USA*

617-432-7421

arbesman@hcp.med.harvard.edu

It has long been known that scientific output proceeds on an exponential increase, or more properly, a logistic growth curve. The interplay between effort and discovery is clear, and the nature of the functional form has been thought to be due to many changes in the scientific process over time. Here I show a quantitative method for examining the ease of scientific progress, another necessary component in understanding scientific discovery. Using examples from three different scientific disciplines – mammalian species, chemical elements, and minor planets – I find the ease of discovery to conform to an exponential decay. In addition, I show how the pace of scientific discovery can be best understood as the outcome of both scientific output and ease of discovery. A quantitative study of the ease of scientific discovery in the aggregate, such as done here, has the potential to provide a great deal of insight into both the nature of future discoveries and the technical processes behind discoveries in science.

# Introduction

Although precise measurement of the process of scientific discovery is difficult (Bettencourt et al. 2008), it is well-known that scientific output, whether measured by scientific papers, number of scientific journals, or even the number of new universities, is considered to be one of exponential growth, or the early stage of a more general logistic function (Price 1986). This is thought to be due to many factors, such as manpower or the bifurcation of disciplines into sub-disciplines (Price 1951). In addition, effort is a factor in this output, and is correlated with that of the sensitivity of the instruments being used. For example, it has been shown that the sensitivity of particle accelerators (Livingston and Blewett 1962) and radio telescopes (Ekers) follows an exponential distribution over time.

I examine three areas where discovery – not simply scientific output – is well-defined and may be easily quantified. These areas of discovery – new mammalian species, chemical elements, and minor planets – also have the added benefit of having proceeded since the Scientific Revolution so may be examined over long time frames. Both the cumulative number of chemical elements and mammalian species (Reeder, Helgen and Wilson 2007) have proceeded along relatively linear relationships with time. Closer examination reveals that these curves might more properly be an escalation of one or more logistic curves ((Price 1986) for elements). Nonetheless, in the aggregate, discovery of mammalian species are characterized at a rate of about 250 new species per decade and chemical elements are discovered at a rate of about 4 per decade.

On the other hand, the cumulative number of minor planets has proceeded along a path much more closely approximated by an exponential growth function (Price 1986), although in the past few years it might be entering the inflection point of a logistic function.

Scientific discovery is dependent on a number of factors, and can be explained in the aggregate by two competing processes: an increase in the rate of scientific output and effort, and a decrease in the ease of scientific discovery. Scientific progress is not simply the outcome of additional money or manpower; there is a clear difficulty (the inverse of ease) to each successive discovery within a discipline. In some clearly described cases, there is even an end to discovery, with the exhaustion of all further discoveries within a discipline. While by no means the rule, an illustrative example is that of the discovery of major internal organs within the human body. This process proceeded from ancient times until the discovery of the paraythyroid gland within humans in 1880 by Ivar Sandström, the last major organ to be discovered (Carney 1996). Understanding the role that ease plays within discovery is important, and must be examined.

## Data and Methods

We must be able to properly quantify the ease with which discoveries are made over time. For each of the three disciplines mentioned earlier, I measured the change in ease of discovery over time, assuming for these fields that the size of a discovery is related to its ease. For mammalian species and minor planets, it is assumed that size is directly proportional to the ease of discovery (larger objects are easier to find, whether on Earth or in space). For chemical elements, it is assumed that size is inversely proportional to ease of discovery, due to the rarity and instability of elements with higher atomic weights.

By combining the year of discovery with the size of discovery, the mean discovery size per year, as proxy for ease of discovery, can be calculated. Each of these curves were fit to exponential curves, conducted using non-linear least squares analyses. Each discipline's data were gathered as described below.

### Minor Planets

The number of minor planets discovered each year was obtained from the Minor Planet Center website, accessed February 13, 2010: http://www.cfa.harvard.edu/iau/lists/NumberedPerYear.html

Using the Minor Planet Center Orbit Database (Minor-Planet-Center 2009), accessed August 18, 2009 (http://www.cfa.harvard.edu/iau/mpc.html) the mean size of minor planets discovered over time were calculated (when the year was missing or unclear, these data were excluded). The absolute magnitude, $H$, is given for each minor planet, and a standard equation is used to calculate an estimated diameter, $D$, in kilometers, where an intermediate value for the albedo, $p_v$, of 0.1 is assumed (Chesley et al. 2002):

$$D = \frac{1329}{p_v} 10^{-0.2H} \qquad (1)$$

### Mammalian Species

The year of description for each mammalian species was derived from *Mammal Species of the World* (Wilson and Reeder 2005), available online: http://www.bucknell.edu/msw3/

The size of mammalian species came from 'Body mass of late Quarternary mammals' (Smith et al. 2003), available online: http://www.esapubs.org/archive/ecol/E084/094/metadata.htm

These two sources were computationally combined, using taxonomic genus and species, for all species described in 1760 or later. Species where the taxonomic names did not match or the sizes were missing were excluded from the ease of discovery calculations. When examining the cumulative number of species over time, all species with date data were included.

### Chemical Elements

The years of discovery for chemical elements were obtained from Thomas Jefferson National Accelerator Facility website (Gagnon 2009), available online (accessed October 27, 2009): http://education.jlab.org/qa/discover_ele.html

The atomic weights of the chemical elements were obtained from 'Atomic weights of the elements 2007 (IUPAC Technical Report)' (Wieser and Berglund 2009), available online (accessed February 11, 2010): http://www.chem.qmul.ac.uk/iupac/AtWt/

## Results

### Empirical Results

In the case of minor planets, the first discovered was Ceres and was large enough to be initially thought a planet. As time proceeded, the number of undiscovered large asteroids diminished, and the mean size of minor planets discovered over time also decreased, where size is measured by diameter in kilometers. In this case, the ease of discovery can be fit by an exponential decay, with a decay rate of approximately 0.0250 ( $R^2 = 0.93$).

For mammalian species, the average physical size of mammals discovered has also decreased over time according to an exponential decay with a decay rate of 0.0258 ( $R^2 = 0.31$), where size is measured by weight in grams. For chemical elements, the ease of discovery over time can be fit by an exponential decay, with a decay rate of 0.00748 ( $R^2 = 0.35$), where size is measured by inverse atomic weight. These curves are all displayed in Figure 1.

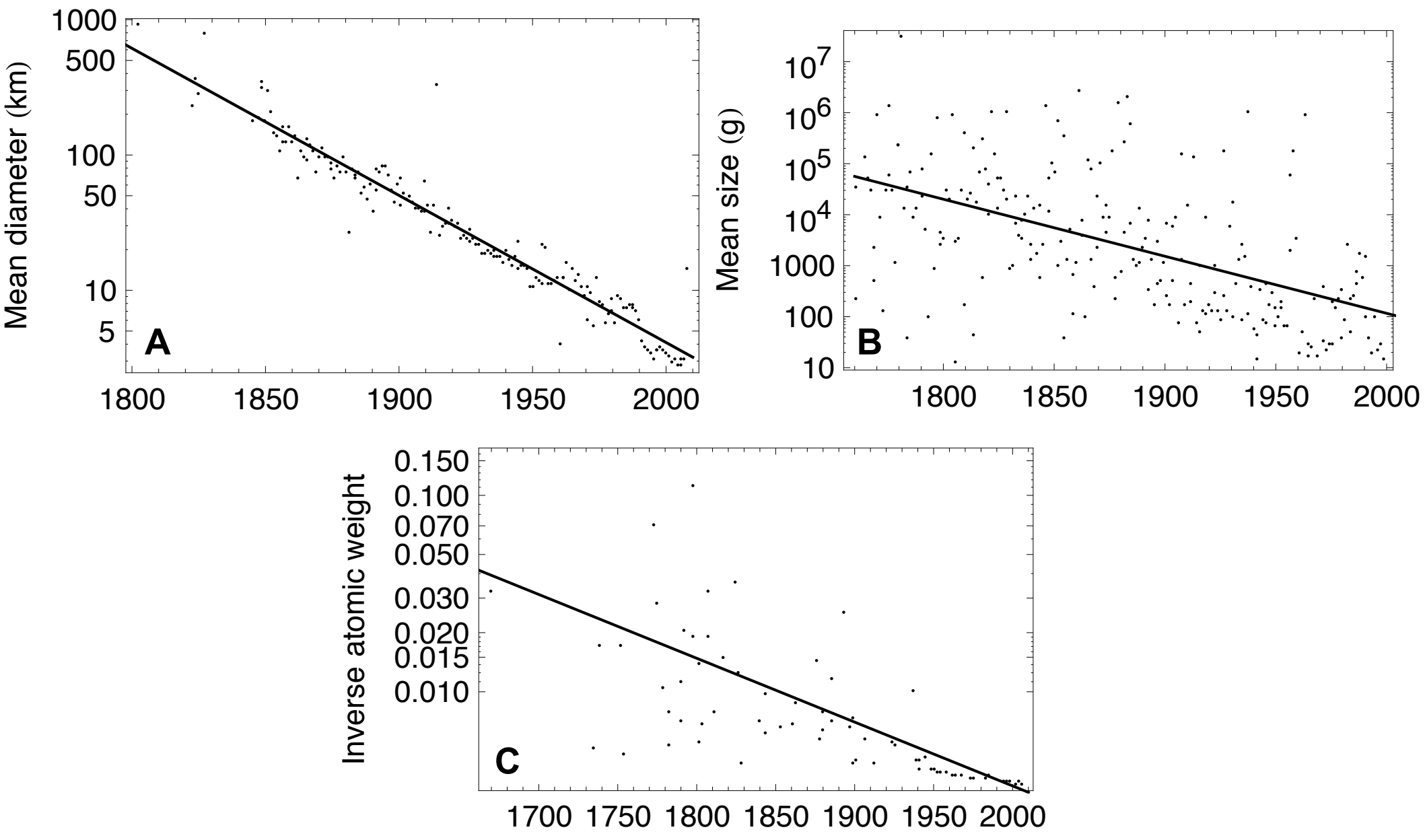

Figure 1. Ease of scientific discovery over time. (A) Mean diameter (kilometers) of minor planets discovered, 1802-2008. (B) Mean physical size (grams) of mammalian species discovered, 1760-2003. (C) Mean inverse of atomic weight of chemical elements discovered, 1669-2006.

## Model

Scientific output, $O(t)$, may be assumed to obey the following functional form, where $r_O$ is the growth rate in scientific output and $O(t)$ follows a logistic curve (Price 1986):

$$O(t) \sim \frac{K}{1 + Ae^{-r_O t}} \qquad (2)$$

$K$ here represents the limiting size of scientific output, and $A$ is a fit constant. More properly, $O(t)$, is often the combination of successive logistic growth functions, such as in the case of the cumulative number of chemical elements (Price 1986), but for simplicity, we assume a single logistic curve.

Guided by the empirical results above, ease of discovery, $E(t)$, obeys the following form:

$$E(t) \sim e^{-r_E t} \qquad (3)$$

Consider the discoveries per unit time, $D(t)$, to be the product of the ease of discovery and the scientific output over time:

$$D(t) = O(t)E(t) \sim \frac{Ke^{-r_E t}}{1 + Ae^{-r_O t}} \qquad (4)$$

The rate of discovery may be thought of as the product of the rate of scientific output and the ease of discovery, where the ease of discovery may be thought of as proportional to the fraction of the output in effort that yields a new discovery.

This results in a function that approximates a logistic function, when $r_E$ is small, because then $e^{-r_E t}$ is approximately 1. As time increases, the functions diverge, but for a suitable range, reasonable fits may be found.

An example fit of the above function for the cumulative number of mammalian species discovered over time (where a single logistic curve is the most appropriate) is shown in Figure 2. This fit was also conducted using a non-linear least squares analysis, where the years were shifted by 1700, for a simpler fit. As can be seen, it fits quite well.

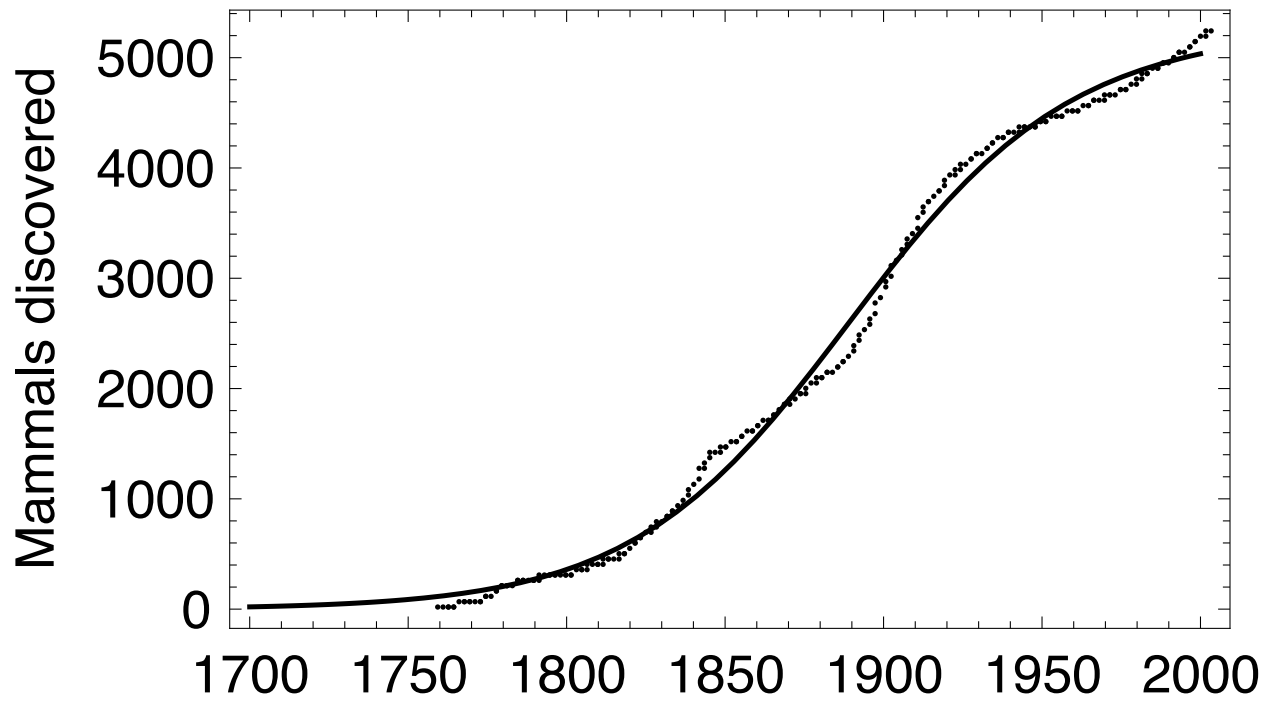

Figure 2. Cumulative number of mammalian species discovered over time. This covers the years 1760-2003. A fit for $D(t)$ was calculated across the mammalian species discovery time series.

## Conclusions

There are many processes involved in scientific progress, and ones that are often related. For example, it can be presumed that breakthroughs in technology play an important role in further discovery, allowing the detection of discoveries of greater difficulty (and lesser ease). Nonetheless, ease in discovery, while clearly important in understanding scientific progress, has not been given much thought or quantitative foundation. A first attempt at quantifying ease's role in scientific discovery is given here. It has been found, in three different areas, to obey a similar exponential decay. While this may be an outcome of the underlying distribution of that which is undiscovered, such as in the case of mammals, where most mammals are small (Pine 1994), these exponential curves provide testable predictions of the future mean sizes of newly discovered mammalian species, minor planets, and chemical elements.

In addition, while precise measurement of the pace of scientific output for each discipline must be made, the above demonstrates that these two competing processes result in an approximately logistic form that fits the cumulative numbers of discoveries within these fields.

Despite all this talk of ease and difficulty though, we should not, as a society or as scientists, find ourselves lapsing into the despondent state of affairs found at the end of the Nineteenth Century that all science was nearly complete (Badash 1972). Nevertheless, it is important to recognize the place of a quantitative study of the ease within scientific discovery in the aggregate, such as done here, in order to better understand the technical processes behind discoveries in science.

Thanks to Nicholas Christakis, Jukka-Pekka Onnela, and Steven Strogatz for reading drafts of this manuscript and providing helpful comments.